\documentstyle[12pt]{article}

\newcommand{\be}{\begin{eqnarray}}
\newcommand{\ee}{\end{eqnarray}}

\newcommand{\pbar}{\bar{p}}

\newcommand{\Qbar}{\bar{Q}}

\newcommand{\xibar}{\bar{\xi}}

\newcommand{\sigmabar}{\bar{\sigma}}
\newcommand{\psibar}{\bar{\psi}}

\newcommand{\calL}{{\cal L}}

\def\pslash{p \kern -0.45em \lower 0.3ex\hbox{/}}
\def\qslash{q \kern -0.45em \hbox{/}}
\def\kslash{k \kern -0.45em \hbox{/}}
\def\Aslash{A \kern -0.45em \hbox{/}}
\def\Bslash{B \kern -0.45em \hbox{/}}
\def\Cslash{C \kern -0.45em \hbox{/}}
\def\Dslash{D \kern -0.45em \hbox{/}}
\def\delslash{\partial \kern -0.45em \hbox{/}}
\def\lradel{\partial \kern -0.7em \lower -2.0ex\hbox{$\leftrightarrow$}}

\title{Hamilton-Jacobi Solution to Soliton Paths
and Triangular Mass Relation in  Two-dimensional
Extended Supersymmetric Theory}
\author{Nobuyuki MOTOYUI, \and Shogo TOMINAGA \and and 
\\Mitsuru YAMADA
\date{August, 2001}
\thanks{E-address: my@mito.ipc.ibaraki.ac.jp}\\
 \\
Department of Mathematical Sciences, Faculty of Sciences,\\
Ibaraki University, Bunkyo 2-1-1, Mito, 310-8512 Japan}

\begin{document}

\maketitle

\begin{abstract}

$D=2,N=2$ generalized Wess-Zumino theory is
investigated by the dimensional reduction from
$D=4,N=1$ theory.
For each solitonic configuration
$(i,j)$ the classical static
solution is solved by the  Hamilton-Jacobi method
of equivalent one-dimensional classical mechanics.
It is easily shown that the Bogomol'nyi mass bound
is saturated by these solutions and triangular mass
inequality
        \[ M_{ij}<M_{ik}+M_{kj}\]
is automatically satisfied.

\end{abstract}

\newpage
Recently, the idea of domain walls in particle field theory is attracting 
much attention in the context of possible extention of the supersymmetry 
algebras \cite{gorsky} \cite{rebhan}. The interest in (1+1)-dimensional 
topological solitons, either supersymmetric or not,has revived because it 
gives approximate solution to the dynamical equation in the space  
direction perpendicular to the wall.

In this note we show that the soliton equation of (1+1)-dimensional 
$(N=2)$-extended generalized Wess-Zumino model can be very simply solved 
by the Hamilton-Jacobi method.
Also we can prove an exact inequality of soliton masses, which shows the 
absolute stability of individual soliton and the existence of  attractive 
force between them.

The two-dimensional $(N=2)$-extended supersymmetric field theroy is most 
easily derived from four-dimensional $(N=1)$ theory through dimensional 
reduction. In  four dimensions, the Lagrangian of supersymmetric 
Wess-Zumino model is given as
        \be {\cal{L}}=\int d^2 \theta d^2 {\theta}^{*} \phi^* \phi +
             \int d^2 \theta W(\phi) + \int d^2 \theta^* W(\phi)^* \ee
        where $\phi$ is a chiral field,
        \be \phi=a+\sqrt 2 \theta^\alpha \psi _\alpha + 
                 \theta^\alpha\theta_\alpha f\ee
        and $W(\phi)$ is the superpotential.
        The current of the supersymmetric charge is
\be j_\mu=\sqrt 2 \left\{\sigma^\rho \sigmabar _\mu \psi \partial_\rho a^* 
         + i\sigma_\mu \psibar W^\prime (a)^* 
\right\} \ee
        where the convention of Weil spinor indices is 
standard~\cite{wb} ~\cite{sohnius}.

By trivial dimensional reduction from $D=4$ to $D=2$,
$(x^\mu)\rightarrow (x^0,0,0,x^1)$,
a Weil spinor $\psi_\alpha (\alpha=1,2)$ is considered as a 
two-component Dirac spinor $\psi=\pmatrix{\psi_1 \cr \psi_2}$,
with identification of 2-dimensional $\gamma$-matrices  in Majorana 
representation
\be \gamma^0&=&\sigma_y \\
\gamma^1&=&-i\sigma_x\\
\gamma_5&=&\gamma^0\gamma^1=-\sigma_z \\
C&=&-\sigma_y\ee
        Then we obtain  the following translation rule of Lorentz scalars :
                \be \xi^\alpha \eta_\alpha&=&-i \xibar^c \eta\ee
where left hand side is a Lorentz invariant product of  Weil spinors 
while right hand side is that of 2-dimensional Dirac spinors¡¥
Note that  $\xibar=\xi^\dagger \gamma_0, \xi^c= C\xibar^T$. 
        Then we have the Lagrangian of $(D=2,N=2)$ Wess-Zumino-type model:
\be \calL=\partial_\mu  a^* \partial^\mu a+i\psibar \gamma^\mu 
\partial_\mu \psi + \frac i2 W^{\prime\prime}(a)\psibar^c \psi 
-\frac i2 W^{\prime\prime}(a)^*\psibar \psi^c -W^\prime (a)W^\prime
 (a)^* \ee
and the current of supersymmetric charge:
\be j_\mu=\sqrt{2}\left\{\gamma^\rho \gamma_\mu \psi \partial_\rho a^* 
- \gamma_\mu \psi^c W^\prime (a)^*\right\}\ee
After elliminating the fermion field by the use of the 
fermion field equation 
\be i\gamma^\mu \partial_\mu \psi-iW^{\prime\prime}(a)^*\psi^c=0\ee
 we have purely bosonic Lagrangian
\be \calL&=&\dot{a}\dot{a}^*-(\nabla a)(\nabla a^*)-|W^\prime (a)|^2 \ee

Let us assume that $W(\phi)$ is a holomorphic function such that 
$W^\prime (a)=0$ has $n$ complex solution
$a_1,a_2,\dots,a_n$.  
Then there are not only $n$ clasical vacuum configurations 
$a(x^0,x^1)=a_\ell, \quad \ell=1,2,\dots,n$ but also at the most
$ n(n-1)$ solitonic cofiguraltions. 
Each of them may be called "$(i,j)$-soliton",
characterized by 
\be a(t,-\infty)&=&a_i\\
a(t,\infty)&=&a_j\ee
 
With  $(i,j)$-soliton in the background, we can check that
the algebra of $(D=2,N=2)$ supersymmetry charge
\be Q=\int_{-\infty}^{\infty}j_0 (x) dx\ee
         undergoes the central extention~\cite{wo}:
\be \left\{Q,\Qbar \right\}&=&2\gamma_\mu P^\mu \\
 \left\{Q,\Qbar ^c \right\}&=& -4 \gamma_5 [W(a_j)^*-W(a_i)^*] \ee
        More explicitly, for center-of-mass frame $(P^\mu)=(M_{ij},0)$
\be \left\{Q_{\alpha},Q^\dagger_{\beta} \right\}&=&2M_{ij}
  \delta_{\alpha\beta} \\
 \left\{Q_\alpha,Q_\beta \right\}&=& -4i (\sigma_x)_{\alpha\beta} 
  \Delta_{ij} W^* \ee
where $\Delta_{ij} W=W(a_j)-W(a_i)$.

From the positivity condition 
\be \left\{ A,A^\dagger \right\} \ge0 \label{eq:20}\ee
with $A=Q_1+ie^{-i\omega}Q_2^\dagger$, we have 
\be M_{ij}\ge 2\Re (e^{-i\omega}\Delta_{ij} W)\ee
Since $\omega$ is arbitrary, 
 we obtain the lower mass bound of $(i,j)$-soliton~\cite{bogomol'nyi},
\be M_{ij}\ge 2 |\Delta_{ij} W|\ee

Actually, this Bogomol'nyi bound is saturated by classical solution. 
To see this, we calculate the static solution of the field equation.
        We start from the Hamiltonian
\be H&=&\dot{a}\dot{a}^*+(\nabla a)(\nabla a^*)+|W^\prime (a)|^2 \ee
where $\nabla a={da}/{dx}$. Writing static solution of $a(t,x)$ 
simply as $a(x)$, we are left with a problem of classical mechanics, 
now $x$ being the  time.
\be {\calL}^\prime&=&(\nabla a)(\nabla a^*)+|W^\prime (a)|^2 \\
H^\prime&=&p_a p_a^* -|W^\prime (a)|^2 \ee
where $p_a$ is the conjugate momentum to $a$.
        The Hamilton-Jacobi equation for the action $S(a,a^*)$
is now
\be \left(\frac{\partial S}{\partial a^*}
\right)
\left(\frac{\partial S}{\partial a}
\right)
 -W^\prime (a)W^\prime (a)^* =E \ee
        For $E=0$ we can write the desired solution
\be S(a,a^*,\alpha)=\alpha W(a)+\frac 1\alpha W(a)^* 
\label{eqn:1}\ee
where the parameter $\alpha = e^{i\beta}$ is a phase.

Another way to the eq.(\ref{eqn:1}) is to change variable from $a$ to $W$.
Since
\be \nabla a=\nabla W \frac{1}{W^\prime (a) }\ee
we see the the Hamiltonian of the  system is
        \be H^\prime =|W^{\prime}|^2\left(
p_W \pbar_W-1
\right)\ee
        For our case of $E=0$ the Hamilton-Jacobi equation is 
\be \left( \frac{\partial S}{\partial W} \right)
\left( \frac{\partial S}{\partial W^*} \right)=1\ee
        And we have the same ation as before.
\be S=\alpha W+\frac{1}{\alpha}W^* \ee
with $ \alpha=e^{i\beta}$.

        In either way the soliton path is given by
\be \frac{\partial S}{\partial \alpha}=W(a)-\frac 1{\alpha^2}W(a)^*
={\rm const.} \ee
Therefore the trajectory $a(x)$ is such that $W(a(x))$is a straight line
in the complex $W$-plane.

The classical mass $M_{ij}$ can be obtained as follows.
\be M_{ij}=\int_{-\infty}^{\infty}{\cal{L}}^\prime dx=2\int _{-\infty}
^{\infty} 
|W^\prime (a(x))|^2 dx  \ee

When $a_\ell$ is a stationary point of the function $W(a)$,
its inverse function $a(W)$ has a branch cut at $W(a_\ell)$.
Usually, each branch cut starts at $W(a_\ell)$ and extends to $\infty$.
Suppose that the segment from $W(a_i)$ to $W(a_j)$ does not cross any of the bra
nch cuts. 
Then using $dx=|da|^2/|dW|$ we can convert the space integral to $W$-integral:
        \be  M_{ij}=2 \left| \int_{W(a_i)}^{W(a_j)}dW \right|
        =2|\Delta_{ij} W|\ee  
 This shows the saturation of the Bogomol'nyi bound.
In passing, we have noticed that the
 $(i,j)$-soliton exists if and only if $W(a_i)$ and $W(a_j)$ 
can be connected by a straight 
line without crossing branch cuts.

Let us further consider the case where three stationary values
$W(a_i),W(a_j)$ and $W(a_k)$ make a  triangle, and no branch cut crosses it. 
Typical example of this is the $Z_N$-symmetric superpotential
$W(\phi)=\frac{1}{N+1}\phi^{N+1}-\phi$.
Then from the  triangular inequality in the $W$-plane, a strict 
mass  relation
\be M_{ik} < M_{ij}+M_{jk}\ee
follows. 
The equality $M_{ik} = M_{ij}+M_{jk}$
would refer to the alignment of three points. 
But in this case the $(i,k)$-soliton does not exist,
since from $a_i$ to $a_k$ it takes twice the infinity of "time".
So this inequality is strict.

The inequality shows the  stability of one-soliton configuration 
together with the existence of attractive force between adjacent solitons.

Now that  the classical saturation of the Bogomol'nyi bound is shown, 
we will  examine the multiplet structure of this case.
Let the argument of $\Delta_{ij}W$ be $\theta$, so that
\be \Delta_{ij}W=e^{i\theta}| \Delta_{ij}W|\ee
        Arranging the supersymmetric generators into $A,B$ 
according to
\be A&=&Q_1+ie^{-i\theta}Q_2^\dagger\\
      B&=&Q_1-ie^{-i\theta}Q_2^\dagger\ee
and substituting the classical static solution $a(x)$,
we can show  the following.

\be      A&=&0\\
        B&=&2\sqrt 2 \int \left\{ W^\prime (a)e^{-i\theta}\psi_1(x)+
i W^\prime (a)^* \psi_2 (x)^* \right\} dx \ee
        The first of them  shows nothing but the bound
(\ref{eq:20}) is saturated for $\omega=\theta$,
while the second further yields the anticommutation relation
\be \left\{B,B^\dagger\right\}=8M_{ij}\ee
        So only $B$ and $B^\dagger$ exite fermionic modes.
Comparing this with the case of no central charge, where 
$A$ and $A^\dagger$ co-operate  with $B$ and $B^\dagger$ 
in  forming supersymmetric irreducible  multiplet, 
we see that the existence of the central charge and the
saturation of the Bogomol'nyi bound caused the multiplet shortening
just like the $D=4,N=2$ case.


\end{document}